\documentclass[aps,pre,onecolumn,superscriptaddress,showkeys]{revtex4-2}

\usepackage[utf8]{inputenc}
\usepackage{float}
\usepackage{amsmath}
\usepackage{amsfonts}
\usepackage{graphicx}
\usepackage{dcolumn}
\usepackage{bm}
\usepackage{hyperref}
\usepackage[mathlines]{lineno}

\begin{document}


\title{Relevant Analytic Spontaneous Magnetization \\ Relation for the Face-Centered-Cubic {I}sing Lattice}


\author{Ba\c{s}er Tamba\c{s}}
\email[]{baser.tambas@gmail.com}
\noaffiliation


\date{\today}

\begin{abstract}
The relevant approximate spontaneous magnetization relations for the simple-cubic and body-centered-cubic Ising lattices have recently been obtained analytically by a novel approach that conflates the Callen--Suzuki identity with a heuristic odd-spin correlation magnetization relation. By exploiting this approach, we study an approximate analytic spontaneous magnetization expression for the face-centered-cubic Ising lattice. We report that the results of the analytic relation obtained in this work are nearly consistent with those derived from the Monte Carlo simulation.
\end{abstract}

\maketitle

\section{\label{sec1}Introduction}
Concerning the phase transition theory~\cite{stanley_1971}, the~Ising model~\cite{ising_1925} is one of the most studied spin systems that exhibits a phase transition at a nonzero finite temperature when the dimension $ d \geq 2 $, resulting in spontaneous magnetization from spontaneously broken discrete $ \mathbb{Z}_{2} $ global symmetry~\cite{smits_2021}. The~Hamiltonian of the Ising model is given by
\[
\mathcal{H}=-J\sum_{\langle i,j \rangle}\sigma_{i}\sigma_{j},
\]
where, $J$ is the coupling strength, $\sum_{\langle i,j \rangle}$ indicates the summation over nearest neighbors, $\sigma_{i}$ may take values $ \pm 1$, and~its symmetry time-reversal $\mathbb{Z}_{2}$ originates in that leaves it invariant under the $\sigma_{i} \rightarrow - \sigma_{i}{}$ transformation. The~exact solutions of one-dimensional and two-dimensional (2D) rectangular lattice Ising models were performed by, respectively, Ising~\cite{ising_1925} and Onsager~\cite{onsager_1944}. The~former has no spontaneous magnetization since it does not undergo a phase transition at a nonzero finite temperature due to its dimensionality. The~latter, besides~having an exact solution, has an exact spontaneous magnetization relation obtained by Yang~\cite{yang_1952}. Though~this was previously obtained by Onsager and Kaufman~\cite{baxter_2012}, they never published their derivation. After~these pioneering contributions by Onsager and Yang for the 2D rectangular lattice Ising model, the~exact solutions and exact spontaneous magnetization relations for the other various 2D lattices, e.g.,~honeycomb, triangular, etc., were also obtained~\cite{houtappel_1950,husimi_1950,syozi_1950,newell_1950,temperley_1950,wannier_1950,potts_1955,naya_1954,potts_1952,lin_1983}. 

The three-dimensional (3D) Ising model, although~subjected to a number of notable attempted solutions~\cite{zhang_2007,wu_2008,perk_2009,zhang_2009,perk_p_2012,zhang_2013,perk_2013} and recent advances~\cite{zhang_2019,suzuki_2021,zhang_2021,zhang_2022}, remains a big mystery as to whether or not it can be solved exactly~\cite{,vis_2022}. The~lack of exact treatments, as~in the case of the 3D Ising model, necessitates the development of approximate methods and concepts in order to explore and detect the emergence properties and critical values in a tractable manner while studying phase transition theory~\cite{domb_1960}. These methods and concepts vary from renormalization group theory~\cite{wilson_1971_1,wilson_1971_2,wilson_1983,goldenfeld_1992}, series expansions~\cite{butera_2002,salman_1998}, field-theoretic~\cite{jasch_2001}, conformal bootstrap~\cite{el_2012,el_2014}, Monte Carlo (MC) simulations~\cite{landau_2014,metropolis_1953,swendsen_1987,wolff_1989,hasenbusch_2001,blote_1996,gupta_1996,murase_2007,lundow_2009,yu_2015,ferrenberg_2018,netz_1992}, and~recently developed machine-learning-aided techniques~\cite{wang_2016,torlai_2016,carrasquilla_2017,wenjian_2017,chung_2021,carleo_2019}.  For~a further discussion on the exact results and approximate methods, we recommend seeing Refs.~\cite{hu_2014,strecka_2015,mccoy_2009,kardar_2007,mccoy_2014,baxter_1982,pelissetto_2002}. The~Ising model, with~its existing exact treatment literature~\cite{ising_1925,onsager_1944,yang_1952,houtappel_1950,husimi_1950,syozi_1950,newell_1950,temperley_1950,wannier_1950,potts_1955,naya_1954,potts_1952,lin_1983,kramers_1941,fisher_1959,kaufman_1949_1,kaufman_1949_2,perk_1980,callen_1963,yang_1988,kac_1952,hurst_1960,montroll_1963}, even if it falls in a limited region, plays a central role in the testing ground for the new methods and techniques and finds interdisciplinary applications in the fields where complexity and intractability emerge, such as economics~\cite{bornholdt_2002,sornette_2006,stauffer_2008}, biology~\cite{weber_2016,matsuda_1981}, sociology~\cite{castellano_2009}, neuroscience~\cite{schneidman_2006,amit_1989}, and~deep learning~\cite{decelle_2021,engel_2001}, etc.

Considering the spontaneous magnetization of the 3D Ising model, a~most recent approach in the context of effective field theory~\cite{strecka_2015} given by Kaya~\cite{kaya_2022_2,kaya_2022_3,kaya_2022_1} has successively led to the derivation of relevant approximate analytic spontaneous magnetization relations for the simple-cubic (SC) and body-centered-cubic Ising lattices. This has been achieved by proposing a heuristic odd-spin correlation magnetization (OSCM) relation by means of associating it to the Callen--Suzuki~\cite{callen_1963,suzuki_1965,suzuki_2002} identity. In~order for the OSCM (heuristic Kaya relation) to apply in this approach, it requires a form of expanded Callen--Suzuki identity in terms of odd-spin correlations. Briefly, the~heuristic OSCM relation implements the idea given in the following statement: the odd-spin correlations in the vicinity of the critical point decay as power-law with the same critical exponent of the spontaneous magnetization but with different amplitudes. This was concluded previously for three-point correlations thereby the exact relations performed by Baxter~\cite{baxter_1975} on the triangular Ising lattice. Such behavior, however, is not very well known for the higher-order odd-correlations when we consider they have relatively restricted literature~\cite{barry_1982}, unlike three-point correlations~\cite{pink_1968,enting_1977,barber_1976,wood_1976,taggart_1982,baxter_1989,kaya_2020,lin_1989,lin_1990}. Nevertheless, the~validity of this approach has been \mbox{verified~\cite{kaya_2022_1,kaya_2022_2}} on the 2D Ising lattices, e.g.,~honeycomb, square, and~triangular, where there are exact expressions for the spontaneous magnetization~\cite{yang_1952,naya_1954,potts_1952} to be compared. For~the SC Ising lattice, it has also been reported that the results derived from this approach~\cite{kaya_2022_2} are in agreement with those of an empirical relation obtained by Talapov and Blöte~\cite{talapov_1996} decades ago. It should also be noted that the critical values needed within this approach, i.e.,~critical temperature and critical exponent, are made supplementary use of the present results in the literature. As~a complementary work, in~the present paper, we shall utilize this approach to obtain a relevant approximate analytic spontaneous magnetization expression for a far more complex crystal structure, the~face-centered-cubic (FCC) Ising lattice, and~perform an MC simulation to compare the results of this analytic relation with those of the MC~simulation. 

It has been pointed out~\cite{newell_1953,zhang_2019,suzuki_2021,zhang_2021,zhang_2022} that in 3D lattices, in~addition to the contribution from the local spin alignment, there is one more type of contribution to the physical properties (including spontaneous magnetization). The~latter is due to the nontrivial topological contributions, i.e.,~the long-range entanglements among the spins. For~a further discussion, see Ref.~\cite{zhang_2019}. We would like to acknowledge that the present method in this paper does not study such nontrivial topological contributions to spontaneous magnetization. 

\section{\label{sec2} Methods and~Results}
\subsection{\label{sec2a}Expansion of the Callen--Suzuki~Identity}
Let us begin by writing down the Callen--Suzuki identity as
\begin{equation}
\label{eq:1}
\left\langle \sigma_{i}\right\rangle = \left\langle \tanh{\left( K \sum_{j}^{q} \sigma_{j} \right) } \right\rangle,
\end{equation}
where, $K = J/(k_{B}T)$, here $T$ and $k_{B}$ correspond, respectively, to~the temperature and the Boltzmann's constant, $\left\langle\cdots\right\rangle$ indicates the ensemble average, $\sigma_{j}$'s are the neighboring spins while $\sigma_{i} $ is the central spin,  and~$q$ denotes the number of nearest neighbors, which is equal to 12 for the FCC lattice. The~$\tanh{}$ term given above can be expanded as
\begin{align}
\label{eq:2}
\tanh{\left( K \sum_{j}^{12} \sigma_{j} \right) } &= A\sum_{(12)} \sigma_{j} +B\sum_{(220)}\sigma_{j_{1}}\sigma_{j_{2}}\sigma_{j_{3}}+C\sum_{(792)}\sigma_{j_{1}}\sigma_{j_{2}}\sigma_{j_{3}}\sigma_{j_{4}}\sigma_{j_{5}}+D\sum_{(792)}\sigma_{j_{1}}\sigma_{j_{2}}\sigma_{j_{3}}\sigma_{j_{4}}\sigma_{j_{5}}\sigma_{j_{6}}\sigma_{j_{7}}\nonumber\\
&\qquad+E\sum_{(220)}\sigma_{j_{1}}\sigma_{j_{2}}\sigma_{j_{3}}\sigma_{j_{4}}\sigma_{j_{5}}\sigma_{j_{6}}\sigma_{j_{7}}\sigma_{j_{8}}\sigma_{j_{9}}+F\sum_{(12)}\sigma_{j_{1}}\sigma_{j_{2}}\sigma_{j_{3}}\sigma_{j_{4}}\sigma_{j_{5}}\sigma_{j_{6}}\sigma_{j_{7}}\sigma_{j_{8}}\sigma_{j_{9}}\sigma_{j_{10}}\sigma_{j_{11}},
\end{align}
where, summations for each odd-spin terms run over the combinations of a given configuration $\sigma_{j_{1}}, \sigma_{j_{2}}, \ldots, \sigma_{j_{12}}$ of the neighboring spins. One can prove the expansion as follows: feeding a fully pointing down configuration, $\{-1,-1,\ldots,-1\}$, into~the $\tanh$ term demonstrates that no terms with an even number of spins are allowed to occur in the right-hand-side because $\tanh$ is an odd function. Moreover, the~permutation symmetry does not allow for more than six different coefficients. These unknown coefficients $A, B, \ldots, F$, can be determined self-consistently by considering all $2^{12}$ configurations of the neighboring spins. Substituting all configurations into Equation~(\ref{eq:2}) results in only six unique equations
\begin{align*}
\tanh\left(2K\right)&=2A-10B+20C-20D+10E-2F, \\
\tanh\left(4K\right)&=4A-12B+8C+8D-12E+4F, \\
\tanh\left(6K\right)&=6A+2B-36C+36D-2E+6F, \\
\tanh\left(8K\right)&=8A+40B-48C-48D+40E+8F, \\
\tanh\left(10K\right)&=10A+110B+132C-132D-100E-10F, \\
\tanh\left(12K\right)&=12A+220B+792C+792D+220E+12F, \\
\end{align*}
through this set of equations, the~coefficients can be obtained straightforwardly as follows
\begin{align*}
A(K)&=\frac{1}{2048}\big[132\tanh{\left(2K\right)}+165\tanh{\left(4K\right)}+110\tanh{\left(6K\right)}+44\tanh{\left(8K\right)}+10\tanh{\left(10K\right)}+\tanh{\left(12K\right)}\big],\nonumber\\
B(K)&=\frac{1}{2048}\big[-36\tanh{\left(2K\right)}-27\tanh{\left(4K\right)}+2\tanh{\left(6K\right)}+12\tanh{\left(8K\right)}+6\tanh{\left(10K\right)}+\tanh{\left(12K\right)}\big],\nonumber\\
C(K)&=\frac{1}{2048}\big[20\tanh{\left(2K\right)}+5\tanh{\left(4K\right)}-10\tanh{\left(6K\right)}-4\tanh{\left(8K\right)}+2\tanh{\left(10K\right)}+\tanh{\left(12K\right)}\big],\nonumber\\
D(K)&=\frac{1}{2048}\big[-20\tanh{\left(2K\right)}+5\tanh{\left(4K\right)}+10\tanh{\left(6K\right)}-4\tanh{\left(8K\right)}-2\tanh{\left(10K\right)}+\tanh{\left(12K\right)}\big],\nonumber\\
E(K)&=\frac{1}{2048}\big[36\tanh{\left(2K\right)}-27\tanh{\left(4K\right)}-2\tanh{\left(6K\right)}+12\tanh{\left(8K\right)}-6\tanh{\left(10K\right)}+\tanh{\left(12K\right)}\big],\nonumber\\
F(K)&=\frac{1}{2048}\big[-132\tanh{\left(2K\right)}+165\tanh{\left(4K\right)}-110\tanh{\left(6K\right)}+44\tanh{\left(8K\right)}-10\tanh{\left(10K\right)}+\tanh{\left(12K\right)}\big].\nonumber\\
\end{align*}

\begin{figure}
\centering 
\includegraphics[trim=250 150 250 115, clip,scale=0.45]{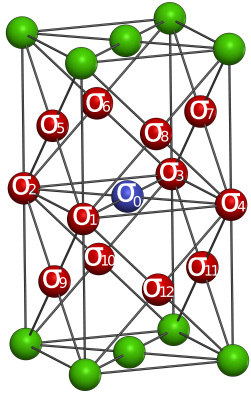}
\caption{(Color Online) \label{fig:1} A diagram of FCC lattice.  $\sigma_{1}, \sigma_{2},\ldots, \sigma_{12}$ (red spins) denote the  neighboring spins whereas $\sigma_{0}$ (blue spin) denotes the central~spin.}
\end{figure}

Substituting Equation~(\ref{eq:2}) into Equation~(\ref{eq:1}) and then taking averages over them yields a form of expanded Callen--Suzuki identity in terms of odd-spin correlations as given~in~the~following
\begin{align}
\label{eq:3}
\left\langle \sigma\right\rangle &= 12A(K) \left\langle \sigma \right\rangle +220B(K) \left\langle \sigma_{j_{1}}\sigma_{j_{2}}\sigma_{j_{3}} \right\rangle +792C(K) \left\langle \sigma_{j_{1}}\sigma_{j_{2}}\sigma_{j_{3}}\sigma_{j_{4}}\sigma_{j_{5}} \right\rangle +792D(K) \left\langle \sigma_{j_{1}}\sigma_{j_{2}}\sigma_{j_{3}}\sigma_{j_{4}}\sigma_{j_{5}}\sigma_{j_{6}}\sigma_{j_{7}} \right\rangle \nonumber\\
&\quad+220E(K) \left\langle \sigma_{j_{1}}\sigma_{j_{2}}\sigma_{j_{3}}\sigma_{j_{4}}\sigma_{j_{5}}\sigma_{j_{6}}\sigma_{j_{7}}\sigma_{j_{8}}\sigma_{j_{9}} \right\rangle +12F(K) \left\langle \sigma_{j_{1}}\sigma_{j_{2}}\sigma_{j_{3}}\sigma_{j_{4}}\sigma_{j_{5}}\sigma_{j_{6}}\sigma_{j_{7}}\sigma_{j_{8}}\sigma_{j_{9}}\sigma_{j_{10}}\sigma_{j_{11}} \right\rangle,
\end{align}
where we have used $\left\langle \sigma_{i} \right\rangle = \left\langle \sigma_{j} \right\rangle \equiv \left\langle \sigma \right\rangle$. The~combinations of the pairwise distances of the neighboring spins on the lattice are different from one another, and~the correlation amplitude between two spins on the different lattice locations is the function of this pairwise distance~\cite{barry_1982}. It will be, therefore, more convenient to group these correlations with respect to the sum of the pairwise distances of the neighboring spins on the lattice for each odd-spin correlation as
\begin{align*}
220 \left\langle \sigma_{j_{1}}\sigma_{j_{2}}\sigma_{j_{3}} \right\rangle &= 8\left\langle \sigma_{1}\sigma_{5}\sigma_{8}  \right\rangle+24\left\langle \sigma_{1}\sigma_{2}\sigma_{5}  \right\rangle+24\left\langle \sigma_{1}\sigma_{5}\sigma_{12}  \right\rangle+48\left\langle \sigma_{1}\sigma_{2}\sigma_{12}  \right\rangle+24\left\langle \sigma_{1}\sigma_{5}\sigma_{10}  \right\rangle+48\left\langle \sigma_{1}\sigma_{3}\sigma_{12}  \right\rangle+12\left\langle \sigma_{1}\sigma_{2}\sigma_{3}  \right\rangle\nonumber\\
&\quad+24\left\langle \sigma_{1}\sigma_{2}\sigma_{11}  \right\rangle+8\left\langle \sigma_{1}\sigma_{6}\sigma_{11}  \right\rangle,\nonumber\\
792 \left\langle \sigma_{j_{1}}\sigma_{j_{2}}\cdots\sigma_{j_{5}} \right\rangle &=  36\left\langle \sigma_{1}\sigma_{2}\sigma_{5}\sigma_{9}\sigma_{12} \right\rangle+48\left\langle \sigma_{1}\sigma_{2}\sigma_{5}\sigma_{8}\sigma_{12} \right\rangle+24\left\langle \sigma_{1}\sigma_{2}\sigma_{4}\sigma_{9}\sigma_{12} \right\rangle+48\left\langle \sigma_{1}\sigma_{2}\sigma_{5}\sigma_{6}\sigma_{12} \right\rangle+24\left\langle \sigma_{1}\sigma_{2}\sigma_{4}\sigma_{8}\sigma_{12} \right\rangle\nonumber\\
&\quad+24\left\langle \sigma_{1}\sigma_{2}\sigma_{5}\sigma_{10}\sigma_{12} \right\rangle+72\left\langle \sigma_{1}\sigma_{2}\sigma_{3}\sigma_{9}\sigma_{12} \right\rangle+96\left\langle \sigma_{1}\sigma_{2}\sigma_{5}\sigma_{11}\sigma_{12} \right\rangle+48\left\langle \sigma_{1}\sigma_{2}\sigma_{3}\sigma_{10}\sigma_{12} \right\rangle+48\left\langle \sigma_{1}\sigma_{4}\sigma_{5}\sigma_{6}\sigma_{9} \right\rangle\nonumber \\
&\quad+72\left\langle \sigma_{1}\sigma_{2}\sigma_{3}\sigma_{11}\sigma_{12} \right\rangle+48\left\langle \sigma_{1}\sigma_{3}\sigma_{5}\sigma_{11}\sigma_{12} \right\rangle+48\left\langle \sigma_{1}\sigma_{3}\sigma_{6}\sigma_{8}\sigma_{12} \right\rangle+48\left\langle \sigma_{1}\sigma_{2}\sigma_{6}\sigma_{11}\sigma_{12} \right\rangle+24\left\langle \sigma_{1}\sigma_{2}\sigma_{3}\sigma_{4}\sigma_{5} \right\rangle\nonumber \\
&\quad+24\left\langle \sigma_{1}\sigma_{2}\sigma_{3}\sigma_{8}\sigma_{12} \right\rangle+36\left\langle \sigma_{1}\sigma_{2}\sigma_{7}\sigma_{10}\sigma_{12} \right\rangle+24\left\langle \sigma_{1}\sigma_{2}\sigma_{3}\sigma_{7}\sigma_{12} \right\rangle,\nonumber\\
792\left \langle \sigma_{j_{1}}\sigma_{j_{2}}\cdots\sigma_{j_{7}} \right \rangle &= 36 \left \langle \sigma_{1}\sigma_{2}\sigma_{3}\sigma_{9}\sigma_{10}\sigma_{11}\sigma_{12} \right\rangle+48\left \langle \sigma_{1}\sigma_{2}\sigma_{3}\sigma_{5}\sigma_{9}\sigma_{10}\sigma_{12} \right\rangle+24\left \langle \sigma_{1}\sigma_{2}\sigma_{3}\sigma_{6}\sigma_{9}\sigma_{10}\sigma_{12} \right\rangle+48\left \langle \sigma_{1}\sigma_{2}\sigma_{4}\sigma_{5}\sigma_{8}\sigma_{11}\sigma_{12} \right\rangle \nonumber\\
&\quad+24\left \langle \sigma_{1}\sigma_{2}\sigma_{3}\sigma_{4}\sigma_{9}\sigma_{10}\sigma_{12} \right\rangle+24\left \langle \sigma_{1}\sigma_{2}\sigma_{3}\sigma_{5}\sigma_{8}\sigma_{9}\sigma_{12} \right\rangle+72\left \langle \sigma_{1}\sigma_{2}\sigma_{3}\sigma_{5}\sigma_{6}\sigma_{8}\sigma_{12} \right\rangle+96\left \langle \sigma_{1}\sigma_{2}\sigma_{3}\sigma_{5}\sigma_{10}\sigma_{11}\sigma_{12} \right\rangle\nonumber\\
&\quad+48\left \langle \sigma_{1}\sigma_{2}\sigma_{5}\sigma_{7}\sigma_{9}\sigma_{11}\sigma_{12} \right\rangle+48\left \langle \sigma_{1}\sigma_{2}\sigma_{3}\sigma_{7}\sigma_{10}\sigma_{11}\sigma_{12} \right\rangle+72\left \langle \sigma_{1}\sigma_{2}\sigma_{3}\sigma_{5}\sigma_{7}\sigma_{9}\sigma_{12} \right\rangle+48\left \langle \sigma_{1}\sigma_{2}\sigma_{3}\sigma_{6}\sigma_{7}\sigma_{9}\sigma_{12} \right\rangle \nonumber\\
&\quad+48\left \langle \sigma_{1}\sigma_{2}\sigma_{3}\sigma_{4}\sigma_{5}\sigma_{11}\sigma_{12} \right\rangle+48\left \langle \sigma_{1}\sigma_{2}\sigma_{3}\sigma_{5}\sigma_{8}\sigma_{11}\sigma_{12} \right\rangle+24\left \langle \sigma_{1}\sigma_{2}\sigma_{3}\sigma_{4}\sigma_{8}\sigma_{10}\sigma_{12} \right\rangle+24\left \langle \sigma_{1}\sigma_{2}\sigma_{3}\sigma_{4}\sigma_{5}\sigma_{7}\sigma_{12} \right\rangle \nonumber\\
&\quad+36\left \langle \sigma_{1}\sigma_{2}\sigma_{3}\sigma_{7}\sigma_{8}\sigma_{11}\sigma_{12} \right\rangle+24\left \langle \sigma_{1}\sigma_{2}\sigma_{3}\sigma_{5}\sigma_{7}\sigma_{11}\sigma_{12} \right\rangle,\nonumber\\
220\left\langle \sigma_{j_{1}}\sigma_{j_{2}}\cdots\sigma_{j_{9}} \right\rangle &= 8\left\langle \sigma_{1}\sigma_{2}\sigma_{3}\sigma_{5}\sigma_{6}\sigma_{9}\sigma_{10}\sigma_{11}\sigma_{12} \right\rangle+24\left\langle \sigma_{1}\sigma_{2}\sigma_{3}\sigma_{4}\sigma_{5}\sigma_{6}\sigma_{7}\sigma_{8}\sigma_{12} \right\rangle+24\left\langle \sigma_{1}\sigma_{2}\sigma_{3}\sigma_{5}\sigma_{8}\sigma_{9}\sigma_{10}\sigma_{11}\sigma_{12} \right\rangle\nonumber\\
&\quad+48\left\langle \sigma_{1}\sigma_{2}\sigma_{3}\sigma_{4}\sigma_{5}\sigma_{6}\sigma_{8}\sigma_{9}\sigma_{12} \right\rangle+24\left\langle \sigma_{1}\sigma_{2}\sigma_{3}\sigma_{5}\sigma_{6}\sigma_{7}\sigma_{8}\sigma_{11}\sigma_{12} \right\rangle+48\left\langle \sigma_{1}\sigma_{3}\sigma_{5}\sigma_{6}\sigma_{7}\sigma_{9}\sigma_{10}\sigma_{11}\sigma_{12} \right\rangle\nonumber\\
&\quad+12\left\langle \sigma_{1}\sigma_{2}\sigma_{3}\sigma_{4}\sigma_{5}\sigma_{7}\sigma_{9}\sigma_{11}\sigma_{12} \right\rangle+24\left\langle \sigma_{1}\sigma_{2}\sigma_{3}\sigma_{4}\sigma_{5}\sigma_{6}\sigma_{7}\sigma_{10}\sigma_{12} \right\rangle+8\left\langle \sigma_{1}\sigma_{2}\sigma_{3}\sigma_{5}\sigma_{7}\sigma_{8}\sigma_{10}\sigma_{11}\sigma_{12} \right\rangle,\nonumber\\
\end{align*}
where, the~correlations specified by the neighboring spins on the lattice, as~indicated in Fig.~\ref{fig:1}, represent the entire group to which they belong, and~the correlations here have been chosen arbitrarily from the group in which they are included.  On~the other hand, the~last term in Equation~(\ref{eq:3}) has not been grouped since the sum of the pairwise distances of all the combinations on the lattice are equal to each~other. \\

For the proof, the~derivation, and~further discussions for the Callen--Suzuki identity and its odd-spin correlation expansion for the six neighboring spins (e.g., SC lattice), similar to what we detailed in this paper, see Refs.~\cite{fisher_1959,perk_p_2012,strecka_2015,suzuki_2002,barry_1982}.
\subsection{\label{sec2b}Derivation of the Analytic Relation for the Spontaneous~Magnetization}
We now proceed by introducing the heuristic Kaya relation as given in Refs.~\cite{kaya_2022_1,kaya_2022_2,kaya_2022_3} in the form
\begin{equation}
\label{eq:4}
    \left\langle \sigma_{j_{1}}\sigma_{j_{2}}\ldots \sigma_{j_{k}} \right\rangle= \lambda_{k,r}\left\langle \sigma \right\rangle + \left(1-\lambda_{k,r} \right)\left\langle \sigma \right\rangle^{\frac{1+\beta}{\beta}},
\end{equation}
in which $k$ on the subscript is an odd number that refers to the order of the odd-correlations, $r$ labels the groups, $\beta$ denotes the critical exponent, and~$\lambda_{k,r}$ may take $0\leq \lambda_{k,r} \leq 1$ values by definition~\cite{kaya_2022_1,kaya_2022_2,kaya_2022_3}. Applying Equation~(\ref{eq:4}) to all the grouped correlations and then substituting these expressions into Equation~(\ref{eq:3}) leads to
\begin{align}
\label{eq:5}
\left\langle\sigma\right\rangle &= \left\langle\sigma\right\rangle \big( 12A+B\mathcal{C}_{3}+C\mathcal{C}_{5}+D\mathcal{C}_{7}+E\mathcal{C}_{9}+F\mathcal{C}_{11} \big) \nonumber\\
&\quad+\left\langle\sigma\right\rangle^{\frac{1+\beta}{\beta}}\big( 220B+792C+792D+220E+12F \nonumber\\
&\qquad\qquad\qquad-B\mathcal{C}_{3}-C\mathcal{C}_{5}-D\mathcal{C}_{7}-E\mathcal{C}_{9}-F \mathcal{C}_{11} \big),\nonumber\\
\end{align}
where,
\begin{align*}
\mathcal{C}_{3} &\equiv 8\lambda_{3,1}+24\lambda_{3,2}+24\lambda_{3,3}+48\lambda_{3,4}+24\lambda_{3,5}+48\lambda_{3,6}+12\lambda_{3,7}+24\lambda_{3,8}+8\lambda_{3,9},\nonumber\\
\mathcal{C}_{5} &\equiv 36\lambda_{5,1}+48\lambda_{5,2}+24\lambda_{5,3}+48\lambda_{5,4}+24\lambda_{5,5}+24\lambda_{5,6}+72\lambda_{5,7}+96\lambda_{5,8}+48\lambda_{5,9}+48\lambda_{5,10}\nonumber\\
&\quad+72\lambda_{5,11}+48\lambda_{5,12}+48\lambda_{5,13}+48\lambda_{5,14}+24\lambda_{5,15}+24\lambda_{5,16}+36\lambda_{5,17}+24\lambda_{5,18}, \nonumber\\
\mathcal{C}_{7} &\equiv 36\lambda_{7,1}+48\lambda_{7,2}+24\lambda_{7,3}+48\lambda_{7,4}+24\lambda_{7,5}+24\lambda_{7,6} +72\lambda_{7,7}+96\lambda_{7,8}+48\lambda_{7,9}+48\lambda_{7,10} \nonumber\\
&\quad+72\lambda_{7,11}+48\lambda_{7,12}+48\lambda_{7,13}+48\lambda_{7,14}+24\lambda_{7,15}+24\lambda_{7,16}+36\lambda_{7,17}+24\lambda_{7,18}, \nonumber\\
\mathcal{C}_{9} &\equiv 8\lambda_{9,1}+24\lambda_{9,2}+24\lambda_{9,3}+48\lambda_{9,4}+24\lambda_{9,5}+48\lambda_{9,6} +12\lambda_{9,7}+24\lambda_{9,8}+8\lambda_{9,9},\nonumber\\
\mathcal{C}_{11} &\equiv 12\lambda_{11,1}.\nonumber
\end{align*}
\begin{figure}
\centering 
\includegraphics[trim=0 0 0 0, clip,scale=0.6]{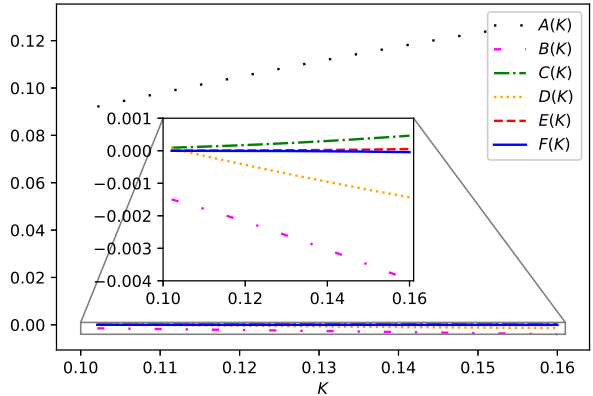}
\caption{(Color Online) \label{fig:2} The plots of $A(K)$, $B(K)$, $C(K)$, $D(K)$, $E(K)$, and~$F(K)$ versus $K$. The~lines loosely-dotted (black), loosely-dash-dotted (magenta), dash-dotted (green), dotted (orange), dashed (red), and~solid (blue) correspond to $A(K)$, $B(K)$, $C(K)$, $D(K)$, $E(K)$, and~$F(K)$, respectively. The~inset zooms in the selected region in the~figure.}
\end{figure}

Now, let us cancel $\left\langle\sigma\right\rangle$ terms in Equation~(\ref{eq:5}), then, it turns out to be
\vspace{2pt}
\begin{align}
\label{eq:5revised}
1 &= \big( 12A+B\mathcal{C}_{3}+C\mathcal{C}_{5}+D\mathcal{C}_{7}+E\mathcal{C}_{9}+F\mathcal{C}_{11} \big) \nonumber\\
&\quad+\left\langle\sigma\right\rangle^{\frac{1}{\beta}}\big( 220B+792C+792D+220E+12F \nonumber\\
&\qquad\qquad\quad-B\mathcal{C}_{3}-C\mathcal{C}_{5}-D\mathcal{C}_{7}-E\mathcal{C}_{9}-F \mathcal{C}_{11} \big).\nonumber\\ 
\end{align}

As these $\lambda_{k,r}$s are unknown, we restrict ourselves to adopting an approximation at this stage. As~can be seen in  Fig.~\ref{fig:2}, $E(K)$ and $F(K)$ give almost no contribution. Thus, we can omit the $\mathcal{C}_{9}$ and $\mathcal{C}_{11}$ terms since they are multiplied by $E(K)$ and $F(K)$, respectively. The~remaining terms $\mathcal{C}_{3}$, $\mathcal{C}_{5}$, and~$\mathcal{C}_{7}$ can be assumed to be nearly equal to each other. Hence, we summarize the approximation we shall adopt as {follows} 
\begin{subequations}
    \label{eq:6}
    \begin{equation}
    \label{eq:6a}
       \mathcal{C}_{3}\simeq\mathcal{C}_{5}\simeq\mathcal{C}_{7}, 
    \end{equation}
    \begin{equation}
    \label{eq:6b}
        E(K),F(K)\approx0.
    \end{equation}   
\end{subequations}

Within this approximation, we reduce the number of unknown parameters to only one, which is $\mathcal{C}_{3}$, and~then calculate this single unknown parameter by using the behavior of the spontaneous magnetization in the vicinity of the critical point $K_{c}$ as given by
\[
    \left\langle\sigma\right\rangle = 
\begin{cases}
    \neq0,&  K>K_{c},\\
    0,              & K \leq K_{c}.
\end{cases}
\]

Once the approximation procedure given in Equations~(\ref{eq:6a}) and (\ref{eq:6b}) is applied, Equation~(\ref{eq:5revised}) then becomes
\begin{align}
1 &= 12A+\mathcal{C}_{3}\big(B(K)+C(K)+D(K)\big) \nonumber\\
&\quad+\left\langle\sigma\right\rangle^{\frac{1}{\beta}}\bigg[220B(K)+792C(K)+792D(K)-\mathcal{C}_{3}\big(B(K)+C(K)+D(K)\big)\bigg].\label{eq:7}
\end{align}

To determine the $\mathcal{C}_{3}$ from Equation~(\ref{eq:7}) with the aid of the aforementioned behavior of spontaneous magnetization, we need the $K_{c}$ value. For~which we take $K_{c}= 0.1270707\pm0.0000002$ as predicted in Ref.~\cite{yu_2015}, which is the most recent result in the literature. There are also some other predictions in  Refs.~\cite{lundow_2009,murase_2007}, nevertheless, all of them are consistent with one another within the respective error bars. From~now on, we abbreviate the terms that contain uncertainty by using shorthand notation, e.g.,~$K_{c}= 0.1270707(2)$, for~simplicity. Now, since the last term $\left\langle\sigma\right\rangle^{\frac{1}{\beta}}$ in Equation~(\ref{eq:7}) vanishes at the critical point $K_{c}$, we thus calculate $\mathcal{C}_{3}$ by
\begin{equation}
\label{eq:8}
    \left(\frac{1-12A(K)}{B(K)+C(K)+D(K)}\right) \bigg|_{K=K_{c}}=\mathcal{C}_{3},
\end{equation}
and we obtain $\mathcal{C}_{3}=74.3922(9)$. Now, substituting the $\mathcal{C}_{3}$ term into Equation~(\ref{eq:7}) and then rearranging it leads to spontaneous magnetization
\begin{equation}
\label{eq:9}
\left\langle\sigma\right\rangle= \left[\frac{1-12A(K)-\mathcal{C}_{3}\big(B(K)+C(K)+D(K)\big) }{220B(K)+792C(K)+792D(K)-\mathcal{C}_{3}\big(B(K)+C(K)+D(K)\big)}\right]^{\beta}.   
\end{equation}

\begin{figure}
\centering 
\includegraphics[trim=0 0 0 0, clip,scale=0.6]{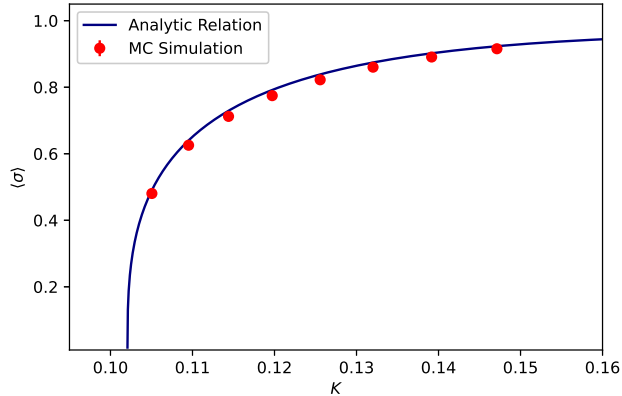}
\caption{(Color Online) \label{fig:3} The plot of $K$ versus $\langle\sigma\rangle$. The~solid (blue) line and the filled points (red) correspond to Equation~(\ref{eq:9}) and the MC data, respectively. The~figure also includes error bars; however, they are invisible because they are smaller than the marker~points.}
\end{figure}

The analytic relation given in Equation~(\ref{eq:9}) has been plotted in  Fig.~\ref{fig:3}, where we have taken the critical exponent $\beta = 0.326 30(22)$ \cite{ferrenberg_2018} for the 3D Ising universality~class.

\subsection{\label{sec2c} Spontaneous Magnetization through MC~Simulation}
To obtain the spontaneous magnetization of the FCC Ising lattice by a different and reliable method, we have also performed an MC simulation by employing the Wolff~\cite{wolff_1989} cluster MC algorithm. It is one of the most efficient algorithms for the 3D Ising model~\cite{binder_2001} because it does not suffer from the critical slowing down in the vicinity of the critical~point. 

In this process, at~each $K$ point, we ran the simulation on a relatively larger $256\times256\times256$ lattice with periodic boundary conditions to avoid the finite size and edge effects. We then collected data at every correlation time interval after running a sufficient number (2000) of thermalization time steps.  To~minimize statistical errors, we set the correlation time interval $\tau$ to 50.  We thus have calculated the spontaneous magnetization using~the~relation 
\[
\left\langle\sigma\right\rangle=\left\langle\left\lvert\frac{1}{N} \sum_{i}\sigma_{i}\right\rvert\right\rangle,
\]
where, $N$ is the number of spins on the lattice, $\lvert\cdots\rvert$ indicates the absolute value, and~averages run over the thermal states (or configurations). The~results have been displayed in Fig.~\ref{fig:3}. The~error bars were also constructed using Jackknife analysis~\cite{landau_2014}; however, they are invisible as they are smaller than the marker~points.

\section{\label{sec3}Discussion and~Conclusions}
In this work, we have studied spontaneous magnetization for the FCC Ising lattice and have obtained an expression as given in Equation~(\ref{eq:9}). To~show the relevance of the analytic relation, we have also performed an MC simulation. The~results of both the MC simulation and Equation~(\ref{eq:9}) have been plotted in Fig.~\ref{fig:3}, where the solid line (blue) and filled marker points (red) correspond to the results of the analytic relation and those derived from the MC simulation, respectively. As~can be seen in Fig.~\ref{fig:3}, there is a minute difference between the centers of the filled points and the solid line. Since the resolution of the figure does not allow for the viewing of error bars,  there is no information as to whether the solid line passes through the error bars of the filled points or not. Consequently, we do not have strong evidence to support the claim that there is an inconsistency between these results. We, therefore, simply interpret that these results are nearly consistent with each other due to the fact that the solid line intersects with the filled points. On~the other hand, if~there is a deviation, this could either arise from the approximation we adopted in the derivation of the analytic relation as given in Equations~(\ref{eq:6a}) and (\ref{eq:6b}) or from the inherent drawbacks of the MC simulations such as finite size effects. Nevertheless, in~the MC simulation, we set the lattice size to $L = 256$, which is a highly sufficient lattice size to minimize this effect as reported in Ref.~\cite{talapov_1996}. Therefore, as~further work, one may clarify these points that we stressed out by calculating the $\mathcal{C}_{3},\mathcal{C}_{5},\ldots,\mathcal{C}_{11}$ values empirically through an MC simulation. This could be simply realized by recalling Equation~(\ref{eq:4}) in the form
\[
\lambda_{k,r}=\frac{\left\langle \sigma_{j_{1}}\sigma_{j_{2}}\ldots \sigma_{j_{k}} \right\rangle - \left\langle \sigma \right\rangle^{\frac{1+\beta}{\beta}}}{\left\langle \sigma \right\rangle - \left\langle \sigma \right\rangle^{\frac{1+\beta}{\beta}}}.
\]

As $K$ is approached to the $K_{c}^{+}$, the~$\left\langle \sigma \right\rangle^{\frac{1+\beta}{\beta}}$ term above goes to zero more rapidly than the other terms. Thus, near~the critical point, the~$\lambda_{k,r}$ simply becomes amplitude ratio, as~shown in the following
\[
\lambda_{k,r}\simeq\frac{\left\langle \sigma_{j_{1}}\sigma_{j_{2}}\ldots \sigma_{j_{k}} \right\rangle}{\left\langle \sigma \right\rangle}.
\]

Once $\lambda_{k,r}$'s terms have been calculated over odd-spin correlation and spontaneous magnetization at the critical point $K_{c}$, it would then be quite straightforward to derive an analytic relation similar to the one performed in this~paper. 

Considering the fact that the spontaneous magnetization expression of the FCC lattice Ising model is a long-standing open problem, we would like to emphasize that the relation we have obtained herein is quite relevant and remarkable regardless of whether it requires a better approximation at the stage of its~derivation. 

\begin{description}
\item[Funding]
This research received no external~funding.

\item[Data Availability]
The datasets generated during and/or analysed during the current study are available from the author on reasonable request.

\item[Acknowledgments]
The author would like to thank Ya\c{s}ar Seher for their support in drawing the FCC~diagram.

\item[Conflict of Iinterest]
The author declares no conflict of~interest.
\end{description}

\bibliography{references}

\end{document}